\documentclass[prd,twocolumn,showpacs]{revtex4}
\usepackage{mathrsfs}
\usepackage{amssymb}
\usepackage{graphicx}
\usepackage[colorlinks,linkcolor=black,anchorcolor=black,citecolor=black]{hyperref}

\def\slashU #1{#1\kern-1.6ex\hbox{\raise0.2ex\hbox{/}}}
\def\slashL #1{#1\kern-1.2ex\hbox{\raise0.0ex\hbox{/}}}

\begin{document}
\title{Intermediate effective interactions and dynamical fermion mass generation of QCD}

\author{Ming-Fan Li}
\email{11006139@zju.edu.cn}
\affiliation{Zhejiang Institute of
Modern Physics, Zhejiang University, Hangzhou, PR China, 310027}

\author{Mingxing Luo}
\email{mingxingluo@zju.edu.cn}
\affiliation{Zhejiang Institute of Modern Physics, Zhejiang
University, Hangzhou, PR China, 310027}

\begin{abstract}
The functional renormalization group equation is expanded to a two-loop form. This two-loop form equation involves one-loop effective action. An intermediate effective action perspective is adopted toward the one-loop effective action. That is to say, the intermediate effective action could not be of the same form of the bare action and one can make an ansatz to it. Thus by focusing on different high dimensional operators, effects of the chosen operators can be investigated. QCD through intermediate fermion-4 interactions is investigated. Of the 6 kinds of fermion-4 interactions generated by one-loop QCD, 4 kinds generate fermion mass while the other 2 kinds degenerate it. The net effect is fermion mass degeneration when dimensionless mass is large. Flow patterns on the $\tilde{m}^2_{\text{phys.}}-\tilde{g}^2$ plane are drawn.
\end{abstract}

\maketitle

\section{Introduction}\label{sec:intro}

Multi-loop quantum corrections are notoriously inaccessible. Complexity of contraction of the Dyson series, complexity inherent to the considered theory, and complexity of the multi-loop momentum integrals make the calculation extremely complicate. So it is meaningful and necessary if there is an approach that can easily capture or analyze multi-loop properties.

Functional renormalization group equation method \cite{Polchinski,Wetterich,Gies-introduction,Rosten-introduction} implements Wilson's idea of `integrating a single momentum slice'. It is exact, nonperturbative, functional and compact. There have been many applications of this method in literature, for example, see \cite{phase-transitions,diffusion,quark-confinement,application-general,gravity,higher-derivative,QED2}, for more applications, see \cite{arxiv_FR}.

The functional renormalization group equation (FRGE) usually cannot be solved exactly, so approximation is necessary. It can be treated perturbatively. Expanded up to two-loop order, it involves one-loop effective action. The calculation from the bare action to the two-loop effective action is complicated. To simplify the calculation, one can make approximations during the procedure.

In this article, we will view the involved one-loop effective action as an intermediate effective action. One can make an ansatz to it. By making different ansatz, different `route' to, or different part of the two-loop effective action can be investigated.

We will take this perspective to analyze the dynamical chiral symmetry breaking of QCD. Lattice calculations, see \cite{AFN,DLP,FHKNS,AFLNS,Hasenfratz,Debbio,LatKMI} for example, show that the critical fermion number for chiral symmetry breaking $N_f^{\ddag}$ is less than the critical fermion number for asymptotic freedom $N_f^{\dag}=33/2$. So there is a conformal window between $N_f^{\ddag}$ and $N_f^{\dag}$. QCD at this situation is asymptotically-free and chirally-symmetric, hence conformal. Lattice calculations indicate that $8<N_f^{\ddag}<13$.

Many theories have been put up to explain the existence of this conformal window. For example, by an ansatz of all-loop $\beta$-function \cite{all-loop-ansatz}, mass-dependent $\beta$-functions \cite{massive-beta-function}, critical scaling laws \cite{scaling-law}, confinement induced gap equation \cite{entropy}, condensation of dynamical chirality \cite{condense}, gap equation through lattice results \cite{lattice_gap}, etc.

In \cite{IRFP}, we investigated this problem through renormalization flows of the fermion mass and the gauge coupling. We displayed that a theory attracted to an IR-attractive fixed-point with a finite fixed dimensionless-mass will not show dynamical fermion mass generation. And the critical fermion number can be determined as the turning point for existence/nonexistence of IR-attractive fixed point. At two-loop order, it is slightly larger than the lower turning point inhabiting in the $\beta$-function of QCD, that is, $51\times 3/19\approx 8$.

However, there we approximated two-loop results with mainly mathematical considerations. In this article, we will approach two-loop corrections through two-loop FRGE with intermediate effective action ansatzed. The central ingredients are the fermion-4 interactions.

It is well-known that fermion-4 interactions are closely related to dynamical chiral symmetry breaking. A lot of work has been done on the Nambu-Jona-Lasinio model or the Thirring theory or other kinds of fermion-4 theories, for example, see \cite{thirring3,Graphene-like,lock,GNJL}. So it is reasonable to expect fermion-4 interactions play an important role in dynamical chiral symmetry breaking of QCD.

The structure of this article is as follows. In section \ref{sec:CalScheme}, we elaborate our calculation scheme; in section \ref{sec:QCDwFF}, we take the procedure for QCD and derive the renormalization equations; in section \ref{sec:mass_generation}, we draw flow patterns; finally, in section \ref{sect:conclusion}, we give our conclusion.

\section{Calculation scheme}\label{sec:CalScheme}

From a bare action $S$, through adding a regulator term,
\begin{equation}\label{}
    \Delta S_k=\frac{1}{2}\int_q\varphi(-q)R_k(q)\varphi(q),
\end{equation}
the functional renormalization group equation (the Wetterich equation) can be derived
\begin{equation}\label{Wetterich-equation-varphi}
    \partial_t\Gamma_k=\frac{1}{2}\text{Tr}[(\partial_tR_k)\tilde{G}_{\varphi\varphi}],
\end{equation}
where $\partial_t=k\partial_k$. $\Gamma_k$ is the scale dependent effective action. $\tilde{G}_{\varphi\varphi}=\langle\varphi\varphi\rangle-\langle\varphi\rangle\langle\varphi\rangle=(\Gamma_k^{(2)}+R_k)^{-1}\equiv\tilde{\Gamma}_k^{(2)-1}$, is the connected two-point Green function of the regulated theory.

The regulator term is to suppress the low energy modes of the theory and let high energy modes intact. Thus during integration of the bare action $S$ to get the effective action $\Gamma$, only high energy modes are integrated.

So there are some conditions on the regulator $R_k$. Firstly, $R_k(q)\sim 0$ when $q^2\gg k^2$. Secondly, $R_k(q)\sim s^2k^2$ for $q^2\ll k^2$. Effectively, the low energy modes gets a mass suppression of mass $s^2k^2$. Thirdly, $R_k(q)\sim 0$ when $k^2\rightarrow 0$, to ensure the regulated theory identical to the original one when the regulator term is removed. For a detailed discussion of the regulator, see \cite{IRFP}.

The parameter $s^2$ measures to what degree the low energy modes are suppressed and quantifies the relative strength of the regulator term to the kinetic term. So to completely exclude the effects of low energy modes (if it is necessary), $s^2$ should be send to infinity after integration.

In literature, there are other kinds of regulators in use and these regulators can also lead to right answers, such as the optimized regulator \cite{optimized}. So there are some intriguing aspects about how to choose regulators. However a thorough discuss of this issue is not the theme of this article. In this article, we use the suppression-parameterized regulator with a sloped-step-function profile, see Fig. \ref{fig:regulator}.

The Eq. (\ref{Wetterich-equation-varphi}) is a functional equation of $\Gamma_k$. It is of one-loop form. It is exact and incorporates all non-perturbative effects. It usually cannot be solved exactly. However, it can be expanded perturbatively, such as
\begin{eqnarray*}
  \partial_t\Gamma_k^{\text{1-loop}} &=& \frac{1}{2}\text{Tr}\Big[(\partial_tR_k)\frac{1}{R_k+S^{(2)}}\Big],  \\
  \partial_t\Gamma_k^{\text{2-loop}} &=& \frac{1}{2}\text{Tr}\Big[(\partial_tR_k)\frac{1}{R_k+(\Gamma^{\text{1-loop}})^{(2)}}\Big].
\end{eqnarray*}
Here $\Gamma_k^{\text{n-loop}}$ indicates the effective actions at n-loop order.

Let $\Delta_n\Gamma$ denote the quantum correction at n-loop order only, then the last equation can be recast into
\begin{equation}\label{eq:FRGE_two_loop}
    \partial_t\Delta_2\Gamma = \frac{1}{2}\text{Tr}[(\Delta_1\Gamma)^{(2)}\partial_t\tilde{G}].
\end{equation}
The one-loop quantum correction is commonly known,
\begin{equation}\label{eq:D1Gamma}
    \Delta_1\Gamma = \frac{1}{2}\text{Tr}\text{ln}\tilde{S}^{(2)}.
\end{equation}

Eq. (\ref{eq:FRGE_two_loop}) is of two-loop form. In principle, it can be used to calculate two-loop renormalization flows. However there are some subtlies in implementation. Nevertheless, it can be separated and used to capture the specified two-loop effects.

In the following, we use the above two equations to discuss the effect of fermion-mass generation/degeneration of intermediate fermion-4 interactions.

At one-loop order, QCD generates the following 6 kinds of fermion-4 intermediate effective interactions (of order $(\partial^2)^0$),
\begin{eqnarray}\label{eq:GammaFF}
  \Delta_1\Gamma_{\text{FF}} &=& \frac{\Delta\lambda_1}{2}\int(\bar{\psi}^{a_f}t^a\psi^{a_f})(\bar{\psi}^{b_f}t^a\psi^{b_f}) \nonumber\\
              &+& \frac{\Delta\lambda_2}{2}\int(\bar{\psi}^{a_f}t^a\gamma^{\mu}\psi^{a_f})(\bar{\psi}^{b_f}t^a\gamma^{\mu}\psi^{b_f}) \nonumber\\
              &+& \frac{\Delta\lambda_3}{2}\int(\bar{\psi}^{a_f}t^a\gamma^{\mu}\gamma^{\nu}\psi^{a_f})(\bar{\psi}^{b_f}t^a\gamma^{\mu}\gamma^{\nu}\psi^{b_f}) \nonumber\\
              &+& \frac{\Delta\lambda_4}{2}\int(\bar{\psi}^{a_f}t^a\gamma^{\mu}\gamma^{\nu}\gamma^{\rho}\psi^{a_f})(\bar{\psi}^{b_f}t^a\gamma^{\mu}\gamma^{\nu}\gamma^{\rho}\psi^{b_f}) \nonumber\\
              &+& \frac{\Delta\lambda_5}{2}\int(\bar{\psi}^{a_f}t^at^b\psi^{a_f})(\bar{\psi}^{b_f}t^at^b\psi^{b_f}) \nonumber\\
              &+& \frac{\Delta\lambda_6}{2}\int(\bar{\psi}^{a_f}t^at^b\gamma^{\mu}\psi^{a_f})(\bar{\psi}^{b_f}t^at^b\gamma^{\mu}\psi^{b_f}).
\end{eqnarray}
Substitute these effective interactions in to Eq. (\ref{eq:FRGE_two_loop}), corrections to fermion mass and gauge coupling can be generated.

\section{QCD through fermion-4 interactions}\label{sec:QCDwFF}

In this section, QCD with intermediate fermion-4 interactions will be considered. The flow equations of fermion mass and gauge coupling will be derived.

We make the following ansatz for the effective action
\begin{eqnarray}\label{action-ansats}
  \Gamma_{\text{QCD}} &=& Z_{\psi}\int_p\bar{\psi}^{a_f}(p)(\slashL{p}+m)\psi^{a_f}(p) \nonumber\\
         &+& \frac{1}{2}Z_A\int_p A^a_{\mu}(-p)\Big(p^2\delta^{\mu\nu}-(1-\frac{1}{\xi})p^{\mu}p^{\nu}\Big)A^a_{\nu}(p) \nonumber\\
         &-& Z_c\int_p \bar{c}^a(p)p^2c^a(p) \nonumber\\
         &+& gZ_{\psi}\sqrt{Z_A}\int_p\int_{p'}\bar{\psi}^{a_f}(p)\gamma^{\mu}t^aA^a_{\mu}(p-p')\psi^{a_f}(p')  \nonumber\\
         &+& gZ_c\sqrt{Z_A}\int_p\int_{p'}f^{abc}(ip_{\mu})\bar{c}^{a}(p)A^b_{\mu}(p-p')c^{c}(p')  \nonumber\\
         &+& gZ_A\sqrt{Z_A}\int_p\int_{p'}\int_q(2\pi)^d\delta^{(d)}(p+p'+q) \nonumber\\
         && ~~~~~~~~ \cdot f^{abc}(-iq_{\mu})A^a_{\nu}(q)A^b_{\mu}(p)A^c_{\nu}(p')  \nonumber\\
         &+& \frac{1}{4}g^2Z_A^2\int_p\int_{p'}\int_q\int_{q'}(2\pi)^d\delta^{(d)}(p+p'+q+q') \nonumber\\
         && ~~~~~~~~ \cdot f^{abe}f^{cde}A^a_{\kappa}(p)A^b_{\lambda}(p')A^c_{\kappa}(q)A^d_{\lambda}(q').
\end{eqnarray}

The gauge $\xi=1$ is used. For simplicity, fermions are assumed to have the same mass. We have used the convention $\{\gamma^{\mu},\gamma^{\nu}\}=-2\delta^{\mu\nu}I_{4\times4}$. This definition of $\gamma$ matrices differs from the convention $\{\gamma^{\mu},\gamma^{\nu}\}=2\delta^{\mu\nu}I_{4\times4}$ by a factor $i$. So in the above expression, the fermion mass is the Wick-rotated mass $m=i\cdot m_{\text{physical}}$.

The regulator terms are
\begin{eqnarray}
  \Delta S_{\psi} &=& \int_p\int_q\bar{\psi}^{a_f}(p)\hat{R}_{\psi}^{a_fb_f}(p,q)\psi^{b_f}(q); \\
  \Delta S_{A} &=& \frac{1}{2}\int_p\int_q A^a_{\mu}(-p)\hat{R}_A^{ab,\mu\nu}(p,q)A^b_{\nu}(q); \\
  \Delta S_{c} &=& \int_p\int_q \bar{c}^a(p)\hat{R}^{ab}_c(p,q)c^b(q),
\end{eqnarray}
with
\begin{eqnarray}
  \hat{R}_{\psi}^{a_fb_f}(p,q) &=& Z_{\psi}\delta^{a_fb_f}\delta_{pq}r_{\psi}(p^2/k^2;s^2)\slashL{p}, \\
  \hat{R}_A^{ab,\mu\nu}(p,q) &=& Z_A\delta^{ab}\delta^{\mu\nu}\delta_{pq}r_A(p^2/k^2;s^2)p^2,\\
  \hat{R}_c^{ab}(p,q) &=& -Z_c\delta^{ab}\delta_{pq}r_A(p^2/k^2;s^2)p^2,
\end{eqnarray}
with $(1+r_{\psi})^2=(1+r_A)$.

The functional renormalization group equation can be derived as
\begin{eqnarray}\label{wetterich-eq}
    \partial_t\Gamma &=& \frac{1}{2}\text{Tr}[(\partial_t\hat{R}_A)\tilde{G}_{AA}]-\text{Tr}[(\partial_t\hat{R}_{\psi})\tilde{G}_{\psi\bar{\psi}}] \nonumber\\
                     &-& \text{Tr}[(\partial_t\hat{R}_c)\tilde{G}_{c\bar{c}}],
\end{eqnarray}
where $\tilde{G}_{AA}$, $\tilde{G}_{\psi\bar{\psi}}$ and
$\tilde{G}_{c\bar{c}}$ are connected Green functions.

Since $\tilde{G}\cdot\tilde{\Gamma}^{(2)}=1$, $\tilde{G}$ can be expressed by inverting $\tilde{\Gamma}^{(2)}$. To derive the flow equations, it can be written that
\begin{eqnarray}
  \tilde{G}_{c\bar{c}} &\approx& \tilde{\Gamma}_{\bar{c}c}^{(2)-1}, \nonumber\\
  \tilde{G}_{AA} &\approx& \Big[\tilde{\Gamma}_{AA}^{(2)}-\tilde{\Gamma}^{(2)}_{A\psi}\tilde{\Gamma}^{(2)-1}_{\bar{\psi}\psi}\tilde{\Gamma}^{(2)}_{\bar{\psi}A}-\tilde{\Gamma}^{(2)}_{A\bar{\psi}}\tilde{\Gamma}^{(2)-1}_{\psi\bar{\psi}}\tilde{\Gamma}^{(2)}_{\psi A}\Big]^{-1}, \nonumber\\
  \tilde{G}_{\psi\bar{\psi}} &\approx& \Big[\tilde{\Gamma}_{\bar{\psi}\psi}^{(2)}-\tilde{\Gamma}^{(2)}_{\bar{\psi}A}(\tilde{\Gamma}^{(2)}_{AA}-\tilde{\Gamma}^{(2)}_{A\bar{\psi}}\tilde{\Gamma}^{(2)-1}_{\psi\bar{\psi}}\tilde{\Gamma}^{(2)}_{\psi A})^{-1}\tilde{\Gamma}^{(2)}_{A\psi}\Big]^{-1}. \nonumber
\end{eqnarray}

The one-loop QCD part has been calculated in \cite{IRFP}. Remained is the fermion-4 part. After calculation, we arrive at the following flow equations.
\begin{eqnarray}
  \frac{\partial_t \tilde{m}^2}{2\tilde{m}^2}+1 &=& -\tilde{g}^2C_2(r)K_m+\frac{k^{d-2}\Delta\lambda_i}{Z_f^2}H_i, \label{flow-of-m} \\
  \frac{\partial_t \tilde{g}^2}{2\tilde{g}^2}+[g] &=& -\tilde{g}^2\big[C_2(r)K_1+C_2(G)K_2-N_fC(r)K_3\big] \nonumber\\
                   &+& \frac{k^{d-2}\Delta\lambda_i}{Z_f^2}T_i. \label{flow-of-g}
\end{eqnarray}
The dimensionless quantities are defined as $\tilde{m}=m/k$, $\tilde{g}^2=[\int d\Omega_d/(2\pi)^d]g^2/k^{4-d}$. And
\begin{eqnarray*}
  K_m &=& d\big[J(1,1,4;\tilde{m}^2)+J(0,2,2;\tilde{m}^2)\big] \nonumber\\
      &-& \frac{(d-1)(d-2)}{d}J(1,1,3;\tilde{m}^2), \\
  K_1 &=& (d-2)\big[J(1,1,4;\tilde{m}^2)+J(0,2,2;\tilde{m}^2)\big] \nonumber\\
      &-& \frac{4(d-2)}{d}J(-1,3,0;\tilde{m}^2)-\frac{(d-1)(d-2)}{d}J(1,1,3;\tilde{m}^2), \\
  K_2 &=& -\frac{d-2}{2}\big[J(1,1,4;\tilde{m}^2)+J(0,2,2;\tilde{m}^2)\big] \nonumber\\
      &+& \frac{2(d-2)}{d}J(-1,3,0;\tilde{m}^2) \nonumber\\
      &+& \frac{3(d-1)}{d}J(1,1,5;\tilde{m}^2)+\frac{2(d-1)}{d}J(0,2,3;\tilde{m}^2) \nonumber\\
      &-& \bigg[\frac{16(d-2)}{d(d+2)}+\frac{d-14}{2}+\frac{8}{d}\bigg]J(2,0,6;\tilde{m}^2), \\
  K_3 &=& -\frac{8}{d}J(-1,3,-2;\tilde{m}^2)+\frac{16(d+4)}{d(d+2)}J(-2,4,-4;\tilde{m}^2) \nonumber\\
      &-& \frac{64}{d(d+2)}J(-3,5,-6;\tilde{m}^2).
\end{eqnarray*}
\begin{eqnarray*}
  \frac{k^{d-2}\Delta\lambda_1}{Z_f^2} &=& -(\tilde{g}^2)^2J_N(2,2,4;\tilde{m}^2)\tilde{m}^2\cdot dC_2(G), \\
  \frac{k^{d-2}\Delta\lambda_5}{Z_f^2} &=& -(\tilde{g}^2)^2J_N(2,2,4;\tilde{m}^2)\tilde{m}^2\cdot 2d, \\
  \frac{k^{d-2}\Delta\lambda_3}{Z_f^2} &=& -(\tilde{g}^2)^2J_N(2,2,4;\tilde{m}^2)\tilde{m}^2\cdot \frac{-1}{2}C_2(G), \\
  \frac{k^{d-2}\Delta\lambda_2}{Z_f^2} &=& -(\tilde{g}^2)^2J_N(1,2,2;\tilde{m}^2)\cdot\frac{3d-2}{d}C_2(G), \\
  \frac{k^{d-2}\Delta\lambda_6}{Z_f^2} &=& -(\tilde{g}^2)^2J_N(1,2,2;\tilde{m}^2)\cdot\frac{2(3d-2)}{d}, \\
  \frac{k^{d-2}\Delta\lambda_4}{Z_f^2} &=& -(\tilde{g}^2)^2J_N(1,2,2;\tilde{m}^2)\cdot\frac{-1}{2d}C_2(G).
\end{eqnarray*}
\begin{eqnarray*}
  H_1 &=& J(-1,2,0;\tilde{m}^2)\cdot C_2(r)\cdot(-1), \\
  H_2 &=& J(-1,2,0;\tilde{m}^2)\cdot C_2(r)d, \\
  H_3 &=& J(-1,2,0;\tilde{m}^2)\cdot C_2(r)(d-2)d, \\
  H_4 &=& J(-1,2,0;\tilde{m}^2)\cdot C_2(r)[2d^2-d(d-2)^2], \\
  H_5 &=& J(-1,2,0;\tilde{m}^2)\cdot N_fC(r)C_2(r)\cdot 4 \\
      &-& J(-1,2,0;\tilde{m}^2)\cdot [C_2(r)-\frac{1}{2}C_2(G)]C_2(r), \\
  H_6 &=& J(-1,2,0;\tilde{m}^2)\cdot [C_2(r)-\frac{1}{2}C_2(G)]C_2(r)d.
\end{eqnarray*}
\begin{eqnarray*}
  T_1 &=& A\cdot [C_2(r)-\frac{1}{2}C_2(G)], \\
  T_2 &=& A\cdot [C_2(r)-\frac{1}{2}C_2(G)](d-2) \\
      &-& A\cdot N_fC(r)(-4), \\
  T_3 &=& A\cdot [C_2(r)-\frac{1}{2}C_2(G)][2d-(d-2)^2], \\
  T_4 &=& A\cdot [C_2(r)-\frac{1}{2}C_2(G)](d-2)(10d-8-d^2) \\
      &-& A\cdot N_fC(r)(-4)(3d-2), \\
  T_5 &=& A\cdot [C_2(r)-\frac{1}{2}C_2(G)][C_2(r)-C_2(G)], \\
  T_6 &=& A\cdot [C_2(r)-\frac{1}{2}C_2(G)][C_2(r)-C_2(G)](d-2) \\
      &-& A\cdot N_fC(r)C_2(G).
\end{eqnarray*}
\begin{eqnarray*}
  A &\equiv& J(-1,3,0)\cdot\tilde{m}^2\frac{d+2}{d}+J(-2,3,-2)\cdot\frac{d-2}{d}.
\end{eqnarray*}

The dimensionless momentum integral $J(a,b,c;\tilde{m}^2)$ is defined by
\begin{eqnarray}\label{J-definition}
  J(a,b,c;\tilde{m}^2) \equiv \int_l\frac{(\partial_t r_A)\cdot k^{2(a+b)-d}(2\pi)^d/\int d\Omega_d}{(l^2)^a[m^2+l^2(1+r_{\psi})^2]^b(1+r_{\psi})^c}.
\end{eqnarray}
 This integral with a suppression-parameterized regulator has been discussed in \cite{IRFP}. Note that here the roles of $b$ and $c$ have been exchanged. When $\tilde{m}^2=0$ and $b+c/2-1>0$,
\begin{equation}\label{}
    J(a,b,c;0)=\frac{1}{b+\frac{c}{2}-1}.
\end{equation}
When $\tilde{m}^2$ is nonzero, it can be evaluated numerically.

\begin{figure}
  \centering
  \includegraphics[width=6.5cm]{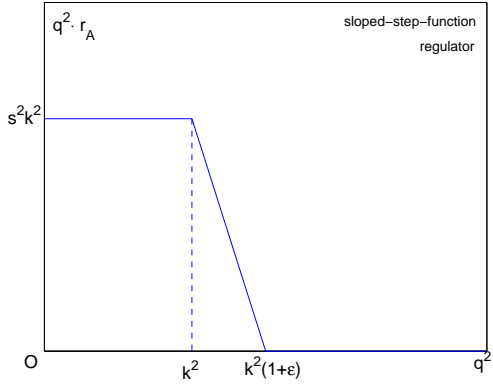}\\
  \caption{Profile of the sloped-step-function regulator.}\label{fig:regulator}
\end{figure}

The regulator we used for numerical calculation is
\begin{equation}\label{eq:regultor_sloped_step}
    l^2\cdot r_A=\frac{s^2}{\epsilon}(k^2+k^2\epsilon-l^2),
\end{equation}
for $l^2\in [k^2,k^2+k^2\epsilon]$; and $l^2\cdot r_A$ be constant elsewhere. Shown in Fig. \ref{fig:regulator} is the profile.

The other momentum integral $J_N(a,b,c;\tilde{m}^2)$ is defined by
\begin{eqnarray}\label{J-definition}
  J_N(a,b,c;\tilde{m}^2) \equiv \int_l\frac{k^{2(a+b)-d}(2\pi)^d/\int d\Omega_d}{(l^2)^a[m^2+l^2(1+r_{\psi})^2]^b(1+r_{\psi})^c}.
\end{eqnarray}
As there is no factor $\partial_t r_A$ in the numerator, one can take limit $s^2\rightarrow\infty$ before integration. When $2(a+b)-d>0$,
\begin{equation}\label{}
    J_N(a,b,c;0)=\frac{1}{2(a+b)-d}.
\end{equation}

\section{Mass generation and degeneration}\label{sec:mass_generation}

In this section, the effects of fermion-4 interactions will be considered. The physical mass $\tilde{m}^2_{\text{physical}}=-\tilde{m}^2$ will be resumed. Renormalization flows in the $\tilde{m}^2_{\text{phys.}}-\tilde{g}^2$ plane will be plotted. In the following, $d=4$ and $N_c=3$.

\begin{figure}
  \centering
  \includegraphics[width=8cm]{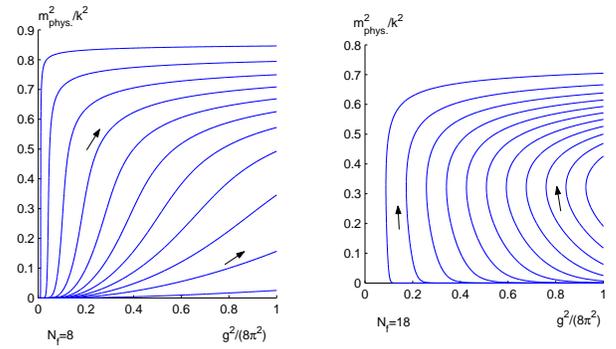}\\
  \caption{Flow patterns of one-loop QCD. Arrows point to infrared.}\label{fig:flow_no_FF}
\end{figure}

Let's first consider the QCD one-loop results without intermediate interactions. The flow patterns for the Wick-rotated mass and gauge coupling have been presented in \cite{IRFP}. Here we draw the flow patterns for the physical mass and gauge coupling. Actually they are the same, respectively. Shown in Fig. \ref{fig:flow_no_FF} are the results. The left panel is typical for $N_f<33/2$, while the right is typical for $N_f>33/2$.

Now Let' s focus on the fermion mass generation/degeneration effect of fermion-4 interactions. From the flow equation (\ref{flow-of-m}), one sees that only $\Delta\lambda_5$ and $\Delta\lambda_6$ contribute negatively to the fermion mass in the infrared indirection. These two interactions lead to fermion mass degeneration.

\begin{figure}
  \centering
  \includegraphics[width=7cm]{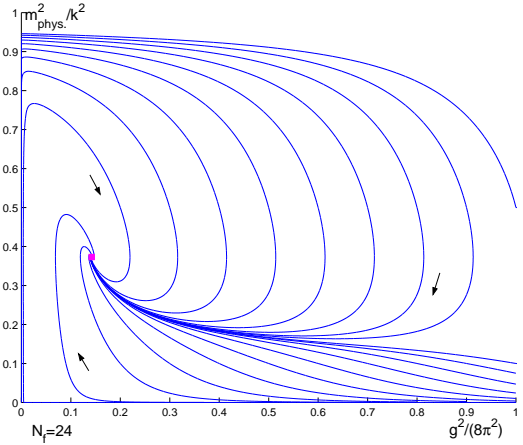}\\
  \caption{Flow pattern for $N_f=24$. Arrows point to infrared.}\label{fig:flow_24_no_Ti}
\end{figure}

\begin{figure}
  \centering
  \includegraphics[width=7cm]{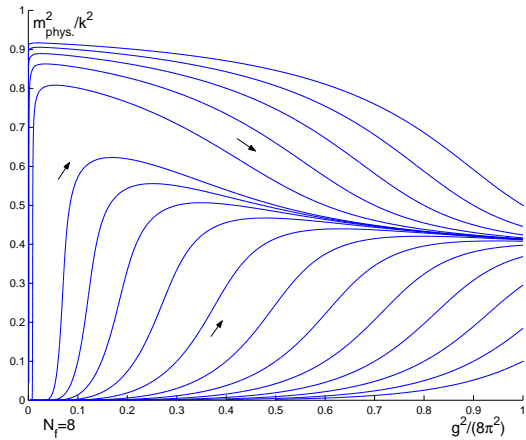}\\
  \caption{Flow pattern for $N_f=8$. Arrows point to infrared.}\label{fig:flow_8_no_Ti}
\end{figure}

Shown in Fig. \ref{fig:flow_24_no_Ti} and Fig. \ref{fig:flow_8_no_Ti} are the flow patterns of one-loop QCD together with the effects of all the 6 kinds of intermediate interactions.

Fig. \ref{fig:flow_24_no_Ti} is typical for $N_f>33/2$. One sees that all the flows go an IR-attractive fixed-point. At this point, $\tilde{m}^2_{\text{phys.}}$ is finite, which means $m^2_{\text{phys.}}=k^2\tilde{m}^2_{\text{phys.}}$ is zero.

Fig. \ref{fig:flow_8_no_Ti} is typical for $N_f<33/2$. All the flows eventually go beyond perturbative region.

The conformal window is not captured, since we have only considered one-loop corrections to the gauge coupling. As has been shown in \cite{IRFP}, with two-loop corrections counted in, the conformal window can be opened.

Now let's consider the gauge coupling correction effect of the fermion-4 interaction. At $\tilde{m}^2_{\text{phys.}}=0$, the flow equation (\ref{flow-of-g}) becomes
\begin{eqnarray}\label{eq:two-loop-gs-ms0-im}
  \frac{\partial_t \tilde{g}^2}{2\tilde{g}^2} = -\tilde{g}^2\Big(\frac{11}{2}-\frac{N_f}{3}\Big)-(\tilde{g}^2)^2\cdot\frac{41}{72}.
\end{eqnarray}
Comparing it with the right two-loop equation
\begin{eqnarray}\label{eq:two-loop-gs-ms0-right}
  \frac{\partial_t \tilde{g}^2}{2\tilde{g}^2} = -\tilde{g}^2\Big(\frac{11}{2}-\frac{N_f}{3}\Big)-(\tilde{g}^2)^2\cdot\Big(\frac{51}{2}-\frac{19N_f}{6}\Big),
\end{eqnarray}
one can see where the difference is. In Eq.(\ref{eq:two-loop-gs-ms0-im}), the coefficient of the second term is negative and independent of $N_f$, which means when $\tilde{g}^2$ is large enough, $\tilde{g}^2$ always increases no matter what $N_f$ is. This is in contrast to the real two-loop result of QCD. This is to say that the fermion-4 interactions contribute to gauge coupling increasing (in the infrared direction) and these contributions do not play a dominate role among all the corrections to the gauge coupling.

\begin{figure}
  \centering
  \includegraphics[width=7cm]{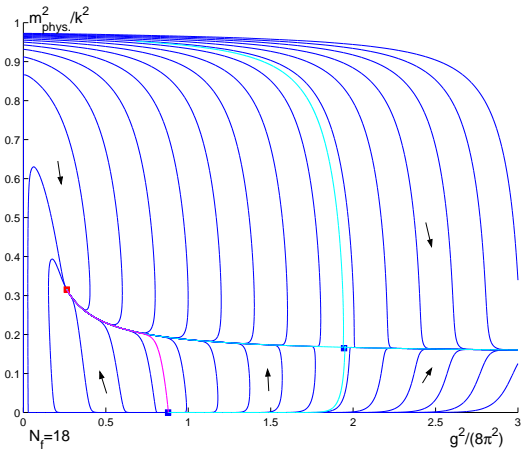}\\
  \caption{Flow pattern of $N_f=18$ with gauge coupling correction of fermion-4 interactions considered. Arrows point to infrared.}\label{fig:flow_18_im_b}
\end{figure}

At this circumstance, the flow pattern for $N_f<33/2$ is the same as in Fig. \ref{fig:flow_8_no_Ti}. When $N_f>33/2$, three non-trivial fixed-points turn up. Shown in Fig. \ref{fig:flow_18_im_b} is the flow pattern.

\section{Conclusion}\label{sect:conclusion}

In conclusion, dynamical fermion mass generation of QCD was investigated through two-loop FRGE with intermediate fermion-4 interactions.

Two-loop quantum corrections are complicated to obtain. The intermediate effective action approach toward the two-loop FRGE provides a convenient tool to analyze and capture two-loop effects.

One-loop QCD generates 6 kinds of fermion-4 interactions. 4 kinds contribute positively to the fermion mass in the IR direction while the other 2 kinds contribute negatively. The net effect is fermion mass degeneration in the IR direction when dimensionless mass is large.

Renormalization flow patterns were drawn on the $\tilde{m}^2_{\text{phys.}}-\tilde{g}^2$ plane. It was displayed that, When chiral symmetry is preserved ($N_f>33/2$), all flows go to an IR-attractive fixed point. When chiral symmetry is broken ($N_f<33/2$), IR-attractive fixed point does not exist. The conformal window was not captured, since fermion-4 interactions do not play a dominate role in all the corrections to the gauge coupling.

\section*{Acknowledgement}

The work is supported in part by the National Science Foundation of
China (10875103, 11135006), National Basic Research Program of China
(2010CB833000).

\begin{appendix}

\section{Some details}\label{app:Details}

$\Gamma^{(2)}$ has the following structure,
\begin{equation}\label{}
    \Gamma^{(2)}=\left(
  \begin{array}{ccccc}
    0 & \Gamma^{(2)}_{\bar{\psi}\psi} & \Gamma^{(2)}_{\bar{\psi}A} & 0 & 0 \\
    \Gamma^{(2)}_{\psi\bar{\psi}} & 0 & \Gamma^{(2)}_{\psi A} & 0 & 0 \\
    \Gamma^{(2)}_{A\bar{\psi}} & \Gamma^{(2)}_{A\psi} & \Gamma^{(2)}_{AA} & \Gamma^{(2)}_{A\bar{c}} & \Gamma^{(2)}_{Ac} \\
    0 & 0 & \Gamma^{(2)}_{\bar{c}A} & 0 & \Gamma^{(2)}_{\bar{c}c} \\
    0 & 0 & \Gamma^{(2)}_{c A} & \Gamma^{(2)}_{c\bar{c}} & 0 \\
  \end{array}
\right).
\end{equation}

With the convention $\{\gamma^{\mu},\gamma^{\nu}\}=-2\delta^{\mu\nu}I_{4\times4}$ and $d$-dimensional Euclidian spacetime,
\begin{equation}\label{}
    \gamma^{\mu}\gamma^{\mu}=(-d)I_{4\times4}.
\end{equation}
\begin{equation}\label{}
    \gamma^{\mu}\gamma^{\rho}\gamma^{\mu}=(d-2)\gamma^{\rho}.
\end{equation}
\begin{equation}\label{}
    \gamma^{\mu}\gamma^{\rho}\gamma^{\nu}+\gamma^{\nu}\gamma^{\rho}\gamma^{\mu}=2\delta^{\mu\nu}\gamma^{\rho}-2\delta^{\rho\nu}\gamma^{\mu}-2\delta^{\rho\mu}\gamma^{\nu}.
\end{equation}

\begin{equation}\label{}
    \slashL{l}\slashL{l}=(-l^2)I_{4\times4}.
\end{equation}

\begin{equation}\label{}
    \text{Tr}(\gamma^{\mu}\gamma^{\nu})=-4\delta^{\mu\nu}.
\end{equation}

Color matrix products:
\begin{equation}\label{}
    t^at^bt^a=[C_2(r)-\frac{1}{2}C_2(G)]t^b.
\end{equation}
\begin{equation}\label{}
    t^at^bt^et^at^b=[C_2(r)-\frac{1}{2}C_2(G)][C_2(r)-C_2(G)]t^e.
\end{equation}
\begin{equation}\label{}
    f^{abc}t^bt^c=\frac{1}{2}iC_2(G)t^a.
\end{equation}

For fundamental representation of $SU(N_c)$,
\begin{equation}\label{}
    t^at^b=\frac{1}{2N_c}\delta^{ab}+\frac{1}{2}(if^{abc}+d^{abc})t^c.
\end{equation}
\begin{equation}\label{}
    t^at^b+t^bt^a=\frac{1}{N_c}\delta^{ab}+d^{abc}t^c.
\end{equation}
\begin{equation}\label{}
    d^{acd}d^{bcd}=\frac{N_c^2-4}{N_c}\delta^{ab}.
\end{equation}
\begin{equation}\label{}
    d^{abc}t^bt^c=[2C_2(r)-\frac{1}{2}C_2(G)-\frac{1}{N_c}]t^a.
\end{equation}
\begin{equation}\label{}
    d^{aac}t^c=0.
\end{equation}
\begin{equation}\label{}
    f^{abe}d^{cde}=0.
\end{equation}
\begin{equation}\label{}
    t^et^at^bt^e=\frac{1}{4N_c}\delta^{ab}C_2(G)+t^at^b[C_2(r)-\frac{1}{2}C_2(G)].
\end{equation}
\begin{equation}\label{}
    \text{Tr}(t^at^bt^c)=\frac{1}{2}if^{abc}C(r).
\end{equation}
\begin{equation}\label{}
    f^{auv}f^{bvw}f^{cwu}=\frac{1}{2}C_2(G)f^{abc}.
\end{equation}

\end{appendix}

\end{document}